\documentstyle[prl,aps,psfig,amsmath]{revtex}
\begin{document}

\draft

\twocolumn [ \hsize\textwidth\columnwidth\hsize\csname
@twocolumnfalse\endcsname

\title{Multichannel $e^{\pm}$ scattering on excited $Ps$ states}

\author{Z.\ Papp${}^{1,2}$, D.\ Caballero${}^{1}$ and C-.Y.\ Hu${}^{1}$}
\address{${}^{1}$ Department of Physics and Astronomy, 
California State University, Long Beach, 90840, California \\
${}^{2}$ Institute of Nuclear Research of the
Hungarian Academy of Sciences, Debrecen, Hungary }
\date{\today}
\maketitle

\begin{abstract}
\noindent
Scattering and reaction cross sections of $e^{\pm}-Ps$ system are calculated
for total angular momentum $L=0, 1$ and $2$ and energies between the
 $Ps(n=2)- Ps(n=3)$
threshold. We solved a set of Faddeev-Merkuriev and Lippmann-Schwinger
integral equations by applying the Coulomb-Sturmian separable expansion
technique. We found that the excited positronium states play dominating roles
in scattering processes.
\end{abstract}

\vspace{0.5cm} 
]

The $e^{\pm}-Ps$ system plays a very important role in studying the antimatter. 
While, on the experimental side, new experiments, 
that involve the positronium in one way or another,
are being carried out or planned \cite{gabrielse}, 
on the theoretical side, the existing calculations are restricted 
for low energy elastic scattering  below the $Ps(n=2)$
threshold (see Ref.\ \cite{igarashi} and references therein). 

Recently, based on a three-potential picture, we have developed a 
new method for treating three-body Coulombic systems \cite{phhky}.
The three-potential formalism results in a set of Faddeev-Merkuriev 
and Lippmann-Schwinger integral equations. These integral equations were 
solved by the Coulomb-Sturmian separable expansion technique.

In this paper we present scattering and reaction 
calculations for energies between the $Ps(n=2)-Ps(n=3)$ threshold and for total 
angular momentum $L=0,1,\ \mbox{and}\ 2$.  First we outline the method of 
Ref.\ \cite{phhky} to the $e^{\pm}-Ps$ system and then present the results.

In the $e^{\pm}-Ps$ system two particles are always identical. 
Let us denote them
by $1$ and $2$, and the non-identical one by $3$.
The  Hamiltonian is given by
\begin{equation}
H=H^0 + v_1^C + v_2^C + v_3^C,
\label{H}
\end{equation}
where $H^0$ is the three-body kinetic energy 
operator and $v_\alpha^C$ denotes the Coulomb
interaction in the subsystem $\alpha$. 
We use the usual
configuration-space Jacobi coordinates
 $x_\alpha$ and $y_\alpha$; $x_\alpha$ is the coordinate
between the pair $(\beta,\gamma)$ and $y_\alpha$ is the
coordinate between the particle $\alpha$ and the center of mass
of the pair $(\beta,\gamma)$.
Thus the potential $v_\alpha^C$, the interaction of the
pair $(\beta,\gamma)$, appears as $v_\alpha^C (x_\alpha)$.
We also use the notation $X=\{x_\alpha,y_\alpha \}\in{\bf R}^6$.

The  Hamiltonian (\ref{H}) is defined in the three-body 
Hilbert space. So, the two-body potential operators are formally
embedded in the three-body Hilbert space,
\begin{equation}
v^C = v^C (x) {\bf 1}_{y},
\label{pot0}
\end{equation}
where ${\bf 1}_{y}$ is a unit operator in the two-body Hilbert space associated
with the $y$ coordinate. The role of a Coulomb potential in a three-body
Coulombic system is twofold. In one hand, it acts like a long-range potential
since it modifies the asymptotic motion. On the other hand, however,
it acts like a short-range potential, since it correlates strongly the 
particles and may support bound states.
Merkuriev introduced a separation of the three-body
configuration space into different
asymptotic regions \cite{fm-book}. 
The two-body asymptotic region $\Omega$ is
defined as a part of the three-body configuration space where
the conditions
\begin{equation}
|x| <  x_0 ( 1  + |y|/ y_0)^{1/\nu},
\label{oma}
\end{equation}
with $x_0, y_0 >0$ and $\nu > 2$, are satisfied.
Merkuriev proposed to split the Coulomb interaction  in 
the three-body configuration space into
short-range and long-range terms 
\begin{equation}
v^C =v^{(s)} +v^{(l)} ,
\label{pot}
\end{equation}
where the superscripts
$s$ and $l$ indicate the short- and long-range
attributes, respectively. 
The splitting is carried out with the help of a splitting function $\zeta$,
\begin{eqnarray}
v^{(s)} (x,y) & = & v^C(x) \zeta (x,y),
\\
v^{(l)} (x,y) & = & v^C(x) [1- \zeta (x,y) ].
\label{potl}
\end{eqnarray}
The function $\zeta$ is defined such that 
\begin{equation}
\zeta(x,y) \xrightarrow{X \to \infty}
\left\{ 
\begin{array}{ll}
1, &  X \in \Omega \\
0  & \mbox{otherwise.}
\end{array}
\right.
\end{equation}
In practice usually the functional form
\begin{equation}
\zeta (x,y) =  
2/\left\{1+ \exp \left[ {(x/x^0)^\nu}/{(1+y/y^0)} \right] \right\},
\label{oma1}
\end{equation}
is used.

In the Hamiltonian (\ref{H}) the potential $v_3^C$, 
acting between the identical
particles, is a repulsive Coulomb
potential which does not support bound states. Consequently, the entire
$v_3^C$ can be considered as long-range potential. Then,
the long-range Hamiltonian is defined as
\begin{equation}
H^{(l)} = H^0 + v_1^{(l)}+ v_2^{(l)}+ v_3^{C},
\label{hl}
\end{equation}
and the three-body Hamiltonian takes the form
\begin{equation}
H = H^{(l)} + v_1^{(s)}+ v_2^{(s)}.
\label{hll}
\end{equation}
So, the Hamiltonian of the $e^{\pm}-Ps$ system appears formally
as a three-body Hamiltonian with two short-range potentials.
The Faddeev procedure is applicable, and, in this case, we 
get a set of two-component Faddeev-Merkuriev integral equations
\begin{eqnarray}
| \psi_1 \rangle &= | \Phi_1^{(l)} \rangle + & G_1^{(l)} v_1^{(s)} 
| \psi_2 \rangle \\
| \psi_2 \rangle &= \phantom{\ | \phi_1 \rangle + } & 
G_2^{(l)} v_2^{(s)} | \psi_1 \rangle,
\label{fm1}
\end{eqnarray}
where $| \phi_\alpha^{(l)}\rangle$ and $G_\alpha^{(l)}$ are eigenstate and
resolvent operator, respectively, of the channel Coulomb Hamiltonian 
\begin{equation}
H_\alpha^{(l)}=H^{(l)}+v_\alpha^{(s)}.
\end{equation}

Since the particles $1$ and $2$ are identical the Faddeev components 
$| \psi_1 \rangle$ and $| \psi_2 \rangle$, in their own natural Jacobi
coordinates, have the same functional forms
\begin{equation}
\langle x_1 y_1 | \psi_1 \rangle = \langle x_2 y_2 | \psi_2 \rangle
= \langle x y | \psi \rangle.
\end{equation}
Therefore we can  determine $| \psi \rangle$ from the first equation
only 
\begin{equation} \label{fmp}
| \psi \rangle = | \Phi_1^{(l)} \rangle +  G_1^{(l)} v_1^{(s)} p {\mathcal P} 
| \psi \rangle,
\end{equation}
where ${\mathcal P}$ is the operator for the permutation of indexes
$1$ and $2$ and $p=\pm 1$ are eigenvalues of ${\mathcal P}$.
We note that although this integral equation has only one component 
yet gives full account 
on the asymptotic and symmetry properties of the system.

We solve this integral equation
by using the Coulomb--Sturmian separable expansion approach.
The Coulomb-Sturmian (CS) functions are defined by
\begin{equation}
\langle r|n l \rangle =\left[ \frac {n!} {(n+2l+1)!} \right]^{1/2}
(2br)^{l+1} \exp(-b r) L_n^{2l+1}(2b r),  \label{basisr}
\end{equation}
with $n$ and $l$ being the radial and
orbital angular momentum quantum numbers, respectively, and $b$ is the size
parameter of the basis.
The CS functions $\{ |n l\rangle \}$
form a biorthonormal
discrete basis in the radial two-body Hilbert space; the biorthogonal
partner defined  by 
$\langle r |\widetilde{n l}\rangle=\langle r |{n l}\rangle/r$. 

Since the three-body Hilbert space is a direct product of two-body
Hilbert spaces an appropriate basis
can be defined as the
angular momentum coupled direct product of the two-body bases 
\begin{equation}
| n \nu  l \lambda \rangle_\alpha =
 | n  l \rangle_\alpha \otimes |
\nu \lambda \rangle_\alpha, \ \ \ \ (n,\nu=0,1,2,\ldots),
\label{cs3}
\end{equation}
where $| n  l \rangle_\alpha$ and $|\nu \lambda \rangle_\alpha$ are associated
with the coordinates $x_\alpha$ and $y_\alpha$, respectively.
With this basis the completeness relation
takes the form (with angular momentum summation implicitly included)
\begin{equation}
{\bf 1} =\lim\limits_{N\to\infty} \sum_{n,\nu=0}^N |
 \widetilde{n \nu l \lambda } \rangle_\alpha \;\mbox{}_\alpha\langle
{n \nu l \lambda} | =
\lim\limits_{N\to\infty} {\bf 1}^{N}_\alpha,
\end{equation}
where $\langle x y | \widetilde{ n \nu l \lambda}\rangle = 
\langle x y | { n \nu l \lambda}\rangle/(x y)$.

We make the following approximation on the 
integral equation (\ref{fmp})
\begin{equation} \label{fmpa}
| \psi \rangle = | \Phi_1^{(l)} \rangle +  G_1^{(l)}
{\bf 1}^{N}_1 v_1^{(s)} p {\mathcal P} {\bf 1}^{N}_1
| \psi \rangle,
\end{equation}
i.e.\ the operator 
$v_1^{(s)} p {\mathcal P}$ is approximated in the three-body
Hilbert space  by a separable form, viz.
\begin{eqnarray}
v_1^{(s)}p {\mathcal P}  & = &
\lim_{N\to\infty} {\bf 1}^{N}_1 v_1^{(s)} p {\mathcal P}  {\bf 1}^{N}_1
\nonumber \\ 
& \approx & {\bf 1}^{N}_1 v_1^{(s)} p {\mathcal P} {\bf 1}^{N}_1 \nonumber \\ 
&  \approx  & \sum_{n,\nu ,n^{\prime },
\nu ^{\prime }=0}^N|\widetilde{n\nu l \lambda}\rangle_1 \;
\underline{v}_1^{(s)}
\;\mbox{}_1 \langle \widetilde{n^{\prime}
\nu ^{\prime} l^{\prime} \lambda^{\prime}}|,  \label{sepfe}
\end{eqnarray}
where $\underline{v}_1^{(s)}=\mbox{}_1 \langle n\nu l \lambda|
v_1^{(s)} p {\mathcal P}  
|n^{\prime }\nu ^{\prime} l^{\prime} \lambda^{\prime}\rangle_1$.
Utilizing the properties of the exchange operator ${\mathcal P}$
these matrix elements can be written in the form $\underline{v}_1^{(s)}= 
p\times \mbox{}_1 \langle n\nu l \lambda| 
v_1^{(s)}|n^{\prime }\nu ^{\prime} l^{\prime} \lambda^{\prime}\rangle_2$, 
and can be evaluated numerically
by using the transformation of the Jacobi coordinates.

Now, with this approximation, the solution of the inhomogeneous
Faddeev-Merkuriev equation
turns into a solution of a matrix equation for the component vector
$\underline{\psi}=
 \mbox{}_1 \langle \widetilde{ n\nu l\lambda} | \psi  \rangle$
\begin{equation}
 \underline{\psi} =  \underline{\Phi}_1^{(l)} + \underline{G}_1^{(l)}  
\underline{v}_1^{(s)}   \underline{\psi}
\label{fn-eq1sm}  
\end{equation}
where 
\begin{equation}
\underline{\Phi}_1^{(l)} = \mbox{}_1 \langle \widetilde{
n\nu l\lambda } |\Phi_1^{(l)} \rangle
\end{equation}
and
\begin{equation}
\underline{G}_1^{(l)}=\mbox{}_1 \langle \widetilde{
n\nu l\lambda} |G_1^{(l)}|\widetilde{n^{\prime}\nu^{\prime} 
l^{\prime} \lambda^{\prime}}\rangle_1.
\end{equation}
The formal solution of Eq.\ (\ref{fn-eq1sm}) is given by
\begin{equation}  \label{fep1}
\lbrack (\underline{G}_1^{(l)})^{-1}-
\underline{v}_1^{(s)}]\underline{\psi }= (\underline{G}_1^{(l)})^{-1}  
\underline{\Phi}_1^{(l)}.
\end{equation}

Unfortunately neither   $\underline{G}_1^{(l)}$ nor  $\underline{\Phi}_1^{(l)}$ 
are known. They are related to the 
Hamiltonian $H_1^{(l)}$, which itself is a complicated three-body Coulomb
Hamiltonian. In the three-potential formalism \cite{phhky}
$\underline{G}_1^{(l)}$ is linked to simpler quantities via solution of a
Lippmann-Schwinger equation,
\begin{equation}
(\underline{G}^{(l)}_1)^{-1}= 
(\underline{\widetilde{G}}_1)^{-1} -
\underline{U}^1,
\label{gleq}
\end{equation}
where 
\begin{equation}
\underline{\widetilde{G}}_{1_{ n \nu l \lambda, 
 n^{\prime}\nu^{\prime}l^{\prime} {\lambda}^{\prime}}} =
 \mbox{}_1\langle \widetilde{n \nu l \lambda} | 
 \widetilde{G}_1 |
 \widetilde{ n^{\prime}\nu^{\prime}l^{\prime}{\lambda}^{\prime}}
 \rangle_1  
 \label{gtilde}
\end{equation}
and 
\begin{equation}
\underline{U}^1_{ n \nu l \lambda,
 n^{\prime}\nu^{\prime} l^{\prime} {\lambda}^{\prime}} =
 \mbox{}_1\langle n\nu l \lambda | U^1 | n^{\prime}\nu^{\prime}
l^{\prime}{\lambda}^{\prime}\rangle_1.
\end{equation}
The operator $\widetilde{G}_1$ is the resolvent operator of the Hamiltonian
\begin{equation} \label{htilde}
\widetilde{H}_1 = H^{0}+v_1^C.
\end{equation}
The polarization potential $U^1$ is defined by
\begin{equation}
U^1=v_2^{(l)}+v_3^C,
\end{equation}
and its matrix elements can again be evaluated numerically.

Similarly, also  $\underline{\Phi}_1^{(l)}$  can be linked
to simpler quantities
\begin{equation}
[ (\underline{\widetilde{G}}_1)^{-1} -
\underline{U}^1]    \underline{\Phi}_1^{(l)} =  
(\underline{\widetilde{G}}_1)^{-1} 
\underline{\widetilde{\Phi}}_1,
\label{eqphil}
\end{equation}
where $
\underline{\widetilde{\Phi}}_{1_{ n \nu l \lambda}}  = 
\mbox{}_1\langle \widetilde{ n \nu l \lambda }| 
\widetilde{\Phi}_1 \rangle$, and $\widetilde{\Phi}_1$ is eigenstate of 
$\widetilde{H}_1$.

The three-particle free Hamiltonian
can be written  as a sum of two-particle free Hamiltonians 
\begin{equation}
H^0=h_{x_1}^0+h_{y_1}^0.
\end{equation}
Consequently the Hamiltonian $\widetilde{H}_1$ of Eq.\ (\ref{htilde}) 
appears as a sum of two Hamiltonians acting on different coordinates 
\begin{equation}
\widetilde{H}_1 =h_{x_1}+h_{y_1},
\end{equation}
with $h_{x_1}=
h_{x_1}^0+v_1^C(x _1)$ and $h_{y_1}=h_{y_1}^0$, which, of course, commute.
Therefore its eigenstate, in CS representation, appears as
\begin{equation}
\mbox{}_1\langle \widetilde{ n \nu l \lambda }| 
\widetilde{\Phi}_1 \rangle = \mbox{}_1\langle \widetilde{ n l}| 
{\phi}_1 \rangle \times \mbox{}_1\langle \widetilde{ \nu \lambda }| 
{\chi}_1 \rangle,
\label{phichi}
\end{equation}
where $|\phi_1 \rangle$ and $|\chi_1 \rangle$ are bound and scattering 
eigenstates of $h_{x_1}$ and $h_{y_1}$, respectively.

The matrix elements of $\widetilde{G}_1$ can be determined by 
making use of the convolution theorem 
\begin{equation}
\widetilde{\underline{G}}_1 (z)=
 \frac 1{2\pi \mathrm{i}}\oint_C
dz^\prime \,\underline{g}_{x_1}(z-z^\prime)\;
\underline{g}_{y_1}(z^\prime),
\label{contourint2}
\end{equation}
where $g_{x_1}$ and $g_{y_1}$ are resolvent operators of 
$h_{x_1}$ and $h_{y_1}$, respectively.
The corresponding CS matrix elements of the two-body Green's operators 
for all complex energies  and of the two-body solutions in Eq.\ (\ref{phichi})
are known analytically.
The contour $C$ should encircle, in positive direction, the 
spectrum of $h_{y_1}$
without penetrating into the spectrum of $h_{x_1}$. 
Further details on the contour and on those  CS matrix elements 
are given in Ref.\ \cite{phhky} and references therein.

The $S$-matrix of the $e^\pm - Ps$ scattering process, 
in the three-potential picture \cite{phhky}, can be decomposed as
\begin{eqnarray}
S_{f i}^{(2)} & = & -2\pi \mbox{i} 
\delta (E_{f}-E_{i}) \nonumber \\
&& \times (\langle 
\widetilde{\Phi }_{1 f}^{(-)}|U^1
|\Phi _{1 i}^{(l)(+)}\rangle +
\langle \Phi _{1 f}^{(l)(-)}|v_1^{(s)} |
\psi _{2 i}^{(+)}\rangle ),
\label{s3}
\end{eqnarray}
where $i$ and $f$ refer to the initial and the final states, respectively.
Having the solutions $\underline{\psi}$ and $\underline{\Phi}^{(l)}$ the
matrix elements can easily be evaluated.

In the numerical calculations we used atomic units. For the parameters
of the splitting function (\ref{oma1}) we took $x_0=10$, $y_0=20$ and
$\nu=2.1$, respectively, and for the size parameter of the CS basis we used
$b=0.2$. In the expansion of the potentials we went up to $11$ angular momentum
channels and, in each angular momentum channels, up to $N=27$ CS functions. 
This way we achieved convergence up to $2-3 \%$ and also the $K$-matrix
were symmetric with a similar accuracy. Some of the results 
were cross-checked by the results of configuration-space differential 
equation calculation \cite{huneni}, and we found again very good agreements.

The results for total angular momentum $L=0$, $L=1$ and $L=2$ are given in 
Tables \ref{tab0}, \ref{tab1} and \ref{tab2}, respectively.
We can see that the excited positronium states play dominating roles
in scattering processes, especially when the total energies approach the
positronium excitation threshold (from above). This behavior is consistent 
with the rather large size of the excited positronium targets, where
the long-range polarization potential play dominant roles. This behavior is
similar to the phenomena found in $\bar{p}-\mbox{Ps}$ multichannel scattering 
process, where this mechanism dominates the antihydrogen
formation cross section (will be published in a separate paper).

This work has been supported by the NSF Grant No.Phy-0088936
and by the OTKA Grants No.\ T026233 and No.\ T029003. We also acknowledge the
generous allocation of computer time at the NPACI, formerly 
San Diego Supercomputing Center,  and at the Department of Aerospace 
Engineering of CSULB.

\begin{table}[tb]
\caption{$L=0$ partial cross sections (in $\pi a_0^2$) 
in the $Ps(n=2)-Ps(n=3)$ gap. Channel numbers $1$,$2$ and 
$3$ denote the
channels $e^\pm(\lambda=0) + Ps(1s)$, $e^\pm(\lambda=0) + Ps(2s)$ and 
$e^\pm(\lambda=1) + Ps(2p)$
respectively.}
\label{tab0}
\begin{tabular}{lcccc}
$k_1$ & Ch.\# & 1 & 2  & 3    \\ \hline
\multicolumn{5}{c}{$L=0\ \ \ p=1$} \\ \hline
      & 1 &  4.384  &  0.043  &  0.032 \\
 0.51 & 2 &  1.125 &  67.79  &  0.245 \\
      & 3 &  0.851  &  0.247 & 301.6 \\ \hline
     & 1 &  4.055 &   0.043  &  0.033 \\
 0.52 & 2 & 0.576  & 60.77 &  45.45  \\
      & 3 &   0.445 &  45.51 &  62.48 \\ \hline
     & 1 &   3.756  &  0.043  &  0.037 \\
 0.53 & 2 &   0.402 &  43.57 &  19.28 \\
      & 3 &   0.349 &  19.32 &  63.77 \\ \hline
      & 1 & 3.484  &  0.044 &   0.041  \\
 0.54 & 2 &   0.324 &  28.39  &  2.12 \\
      & 3 &   0.286  &  2.13 &  71.74 \\ \hline
\multicolumn{5}{c}{$L=0\ \ \ p=-1$} \\ \hline
     & 1 &   12.52 &   0.0001 &   0.0000 \\
 0.51 & 2 &   0.0023 & 121.69 &   0.941 \\
      & 3 &   0.0011 &   0.936 & 263.84  \\ \hline
     & 1 &   11.79  &  0.0001 &   0.0001 \\
 0.52 & 2 &  0.0021&  102.42  & 35.32  \\
      & 3 &    0.0009 &  35.28 &  42.75 \\ \hline
     & 1 &    11.10 &   0.0002 &   0.0002 \\
 0.53 & 2 &    0.0025 &  36.60 &  31.33 \\
      & 3 &    0.0019 &  31.28 &  37.06 \\ \hline
     & 1 &  10.45  &  0.0003  &  0.0004  \\
 0.54 & 2 & 0.0020  &  8.93 & 19.78 \\
      & 3 &   0.0025 &  19.75  & 44.23   
\end{tabular}
\end{table}

\begin{table}[tb]
\caption{$L=1$ partial cross sections (in $\pi a_0^2$) 
in the $Ps(n=2)-Ps(n=3)$ gap. Channel numbers $1$,$2$,$3$ and 
$4$ denote the
channels $e^\pm(\lambda=1) + Ps(1s)$, $e^\pm(\lambda=1) + Ps(2s)$, 
$e^\pm(\lambda=0) + Ps(2p)$ 
and $e^\pm(\lambda=2) + Ps(2p)$, respectively.}
\label{tab1}
\begin{tabular}{lccccc}
$k_1$ & Ch.\# & 1 & 2  & 3    & 4  \\ \hline
\multicolumn{6}{c}{$L=1\ \ \ p=1$} \\ \hline
      & 1 &    20.22  &  0.090  &  0.466 &   0.246   \\
 0.51 & 2 &     2.378 & 296.76  & 30.16 &  29.21   \\
      & 3 &     12.21 &  30.17 &  90.13 & 139.14  \\
      & 4 &     6.49  & 29.40 & 138.90 & 591.54 \\ \hline
      & 1 &   19.24  &  0.086  &  0.617  &  0.317  \\
 0.52 & 2 &   1.16 &  17.76 &  56.03 &  26.56 \\
      & 3 &   8.33 &  55.91 & 127.26  & 66.25  \\
      & 4 &   4.28  & 26.37 &  66.05 & 305.29  \\ \hline
      & 1 &   18.32 &   0.095  &  0.804  &  0.398  \\
 0.53 & 2 &    0.874 &  65.19 &  50.25 &  38.59 \\
      & 3 &  7.47 &  50.16 &  69.26  & 28.60  \\
      & 4 &   3.69  & 38.38 &  28.52 & 174.53   \\  \hline 
      & 1 &    16.91  &  0.207  &  0.916  &  0.599  \\
 0.54 & 2 &    1.48 &  81.55 &  50.21 &  19.94  \\
      & 3 &   6.50  & 50.12  & 66.48  & 10.43  \\
      & 4 &   4.20 &  19.75 &  10.36 & 169.84 \\ \hline
\multicolumn{6}{c}{$L=1\ \ \ p=-1$} \\ \hline
      & 1 &   12.80 &   0.458  &  0.184 &   0.175  \\
 0.51 & 2 &    12.16&  290.22 &  44.19  &  9.62  \\
      & 3 &    4.89 &  44.28 & 737.30 & 135.86 \\
      & 4 &    4.66  &  9.63 & 135.60 & 813.33  \\ \hline
      & 1 &    11.93 &   0.484  &  0.201  &  0.194  \\
 0.52 & 2 &  6.60 &  21.53  &  9.78  & 66.71  \\
      & 3 &   2.71 &   9.77 & 246.08 &  46.32  \\
      & 4 &   2.64 &  66.40  & 46.17 & 308.15  \\ \hline
      & 1 &   11.14 &   0.514 &   0.229  &  0.214 \\
 0.53 & 2 &     4.82 &  81.03  &  9.85 &  79.18 \\
      & 3 &    2.13 &   9.87 &  52.74 &  14.22\\
      & 4 &   2.01   & 78.89  & 14.16  & 94.31 \\ \hline
      & 1 &   10.41   &  0.542 &   0.262 &   0.232  \\
 0.54 & 2 &    3.96  & 145.62 &   4.51 &  50.12  \\
      & 3 &  1.91    & 4.52  &  3.24  &  4.66   \\
      & 4 &   1.68  &  49.97 &   4.65 &  25.25  
 \end{tabular}
\end{table}

\begin{table}[tb]
\caption{$L=2$ partial cross sections (in $\pi a_0^2$) 
in the $Ps(n=2)-Ps(n=3)$ gap. Channel numbers $1$,$2$,$3$ and 
$4$ denote the
channels $e^\pm(\lambda=2) + Ps(1s)$, $e^\pm(\lambda=2) + Ps(2s)$, 
$e^\pm(\lambda=1) + Ps(2p)$ 
and $e^\pm(\lambda=3) + Ps(2p)$, respectively.}
\label{tab2}
\begin{tabular}{lccccc}
$k_1$ & Ch.\# & 1 & 2  & 3    & 4  \\ \hline
\multicolumn{6}{c}{$L=1\ \ \ p=1$} \\ \hline
      & 1 &     7.78  &  0.010  &  0.005  &  0.004   \\
 0.51 & 2 &      0.236 & 917.53 & 174.78 &  64.21  \\
      & 3 &      0.113 & 175.24 & 873.28 & 203.44  \\
      & 4 &      0.097 &  63.72 & 203.03 & 888.39 \\ \hline
      & 1 &     7.58  &  0.025  &  0.004  &  0.012   \\
 0.52 & 2 &       0.333 &  227.40 &  28.16 &  78.54 \\
      & 3 &     0.046  & 28.16  &773.01 &  69.57  \\
      & 4 &     0.153  & 76.55  & 69.05 & 428.25 \\ \hline
      & 1 &     7.38  &  0.048  &  0.003 &   0.025  \\
 0.53 & 2 &     0.438  & 68.33  & 10.13 &  92.31  \\
      & 3 &     0.026  & 10.02  &502.06 &  30.49  \\
      & 4 &     0.217 &  89.59  &  30.02 & 231.33  \\  \hline 
      & 1 &      7.18  &  0.081 &   0.003 &   0.046  \\
 0.54 & 2 &      0.558  & 40.57 &   6.30 &  90.67 \\
      & 3 &       0.028 &   6.29 & 307.42 &  15.42 \\
      & 4 &       0.311  & 88.17 &  15.02 & 127.53  \\ \hline
\multicolumn{6}{c}{$L=1\ \ \ p=-1$} \\ \hline
      & 1 &   6.61  &  0.025  &  0.037  &  0.026  \\
 0.51 & 2 &   0.644 & 941.82 & 194.33 &  36.50  \\
      & 3 &     0.950 & 196.67 &  79.94  & 58.60  \\
      & 4 &    0.656  & 35.57  & 57.09 & 1080.61 \\ \hline
      & 1 &    6.40  &  0.040  &  0.092  &  0.054   \\
 0.52 & 2 &    0.546 & 427.67  &  3.46 &   16.52  \\
      & 3 &     1.22  &  3.29 & 293.60 &   0.754 \\
      & 4 &   0.728  & 15.06  &  0.860 & 510.53  \\ \hline
      & 1 &   6.18  &  0.050  &  0.135 &   0.078   \\
 0.53 & 2 &    0.465 &  159.41 &  37.79 &  20.14  \\
      & 3 &    1.22 &  37.34 & 325.85 &   7.27 \\
      & 4 &    0.711  & 18.57  &  7.22 & 314.93 \\ \hline
      & 1 &     5.95  &  0.055 &   0.221&    0.112  \\
 0.54 & 2 &    0.386 &  72.91 &  49.02  & 22.41 \\
      & 3 &    1.55 &  48.23 & 256.51 &  11.45 \\
      & 4 &    0.757 &  21.33  & 11.44 & 210.85   
\end{tabular}
\end{table}


\begin{references}

\bibitem{gabrielse} G.~Gabrielse, Advances in Atomic, Molecular and
Optical Physics, {\bf 45}, 1 (2001).

\bibitem{igarashi} A.~Igarashi, S.~Nakazaki and A.~Ohsaki,
 Phys.\ Rev.\ A {\bf 61}, 062712 (2000).

\bibitem{phhky} Z.~Papp, C.-Y.~Hu, Z.~T.~Hlousek, B.~K\'onya and
S.~L.~Yakovlev, Phys.~Rev.\ A, (2001).

\bibitem{fm-book}   L.~D.~Faddeev and S.~P.~Merkuriev, {\it Quantum
Scattering Theory for Several Particle Systems} 
(Kluwer, Dordrecht,1993).

\bibitem{huneni} C.-Y.~Hu, J.~Phys.\ B: At.\ Mol.\ Opt.\ Phys.\ {\bf 32},
3077 (1999); C.-Y.~Hu, D.~Caballero and Z.~T.~Hlousek,
ibid, {\bf 34}, 331 (2001).


\end{references}
\end{document}